\documentclass[aps,pra,twocolumn,showpacs,groupedaddress]{revtex4-1}
\usepackage{graphicx}
\usepackage{epstopdf} 
\begin{document}

\title{Lasing versus lasing without inversion in an optically thin gain medium\\ 
near a metal surface}

\author{V.G.~Bordo}
\email{bordo@sdu.dk}

\affiliation{SDU Electrical Engineering, University of Southern Denmark,
Alsion 2, DK-6400 S{\o}nderborg, Denmark}


\date{\today}

\begin{abstract}
A theory of lasing in an optically thin layer of active centers disposed at a metal surface is developed from first principles. The approach is based on a rigorous account of the local field in a close vicinity of a reflective surface which provides a feedback for dipole oscillations in active centers. It is demonstrated that the gain medium thickness plays a crucial role in the lasing condition and controls a switching from conventional lasing to lasing without inversion. The numerical calculations are carried out for erbium doped glass bordering a gold surface where radiation at telecom wavelength (1532 nm) can be generated.
\end{abstract}


\maketitle

\section{Introduction}
The current trend in nanophotonics requires miniaturization of coherent light sources \cite{Odom17,Zhang20}. Nanoscale lasers (nanolasers) are promising for diverse applications, from integration in optical data networks for increasing optical
communication speeds to opening new possibilities in bioimaging and ultra-sensitive chemical analysis.\\
While conventional photonic lasers suffer from the diffraction limit which prevents their reduction in size, the so-called plasmonic nanolasers and spasers \cite{Odom17,Zhang20,Stockman03,Hill07,Noginov09,Oulton09,Toropov21}, which exploit strongly localized electromagnetic field of surface plasmon polaritons (SPPs), allow one to reach the ultimately small dimensions.\\
The active centers which constitute gain medium in nanolasers are confined within a volume with subwavelength dimensions that highlights the quantum electrodynamical effects near an interface or in a cavity \cite{Berman94,Walter06} which are hidden in macroscopic lasers. A molecule emitting light in a close vicinity of a surface undergoes the backaction from it due to the reflected field that leads, in particular, to the oscillating variation of the molecule radiative relaxation time with distance \cite{Drexhage68,Chance74}.\\
This issue dates back to Sommerfeld's paper from 1909, where he calculated the power needed by a dipole antenna above the Earth's surface to radiate radio waves \cite{Sommerfeld09,Sommerfeld64}. The electromagnetic field radiated by the dipole is reflected back to it, so that the forward and backward pathways interfere with each other. As a result, the power needed by the dipole to compensate
the overall losses demonstrates an oscillating behavior as a function of the dipole-surface distance. In this context, both pathways form together a feedback loop which provides either negative or positive feedback from the reflective surface, depending on the dipole position. The same mechanism can be described in terms of the photonic mode density of the dipole surroundings which greatly affects its decay rate \cite{Barnes98}.\\
In a large ensemble of dipoles near a surface the backaction scales with their number and for large enough numbers the positive feedback can prevail over losses in the ensemble \cite{Bordo16}. This implies the existence of a threshold value of the dipoles number density above which the electromagnetic field generation (lasing) becomes possible. The backaction in such a case is averaged over the ensemble and depends on the gain medium location and size. A careful account of this effect is crucial for a proper analysis of the nanolaser and spaser dynamics. It is based on the dyadic Green's function approach and was performed before for nanowire lasers \cite{Bordo13} and core-shell nanoparticle spasers \cite{Pustovit16,Cuerda16,Shahbazyan17,Bordo17a}. \\
In the present paper, we report on a dramatic impact which the backaction has on the laser operation in a simple model system consisting of a thin film of active centers located near a metal surface. Such a structure has been proven to provide a promising route for SPP integrated devices, amplifiers and lasers \cite{Zhang08,Dereux09}. We demonstrate that the lasing condition is determined not just by the population inversion in gain medium as in conventional lasers, but also by a factor which stems from the backaction of the metal surface. As a result, the structure can lase either with or without population inversion, depending on the gain medium thickness.\\ 
The phenomenon of lasing without inversion considered here is based on a phase-sensitive backaction from reflective surfaces. It was predicted before for quite diverse systems: semiconductor nanowire lasers \cite{Bordo13}, SPPs between two metal surfaces \cite{Bordo16} and Rydberg atoms in a beam propagating near a metal surface \cite{Bordo17}. Its underlying mechanism is principally different from all kinds of lasing without inversion which occur in atomic systems in free space \cite{Lu90,Mandel90,Mompart00,Scully13}.\\
The paper is organized as follows. Section \ref{sec:model} introduces the theoretical model which is used to describe the system and calculate the polarization in gain medium. In Sec. \ref{sec:lasing} we derive the lasing condition which is numerically analyzed in Sec. \ref{sec:numerical} for erbium doped glass at a gold surface. The main results of the paper are summarized in Sec. \ref{sec:conclusion}.
\section{Theoretical model}\label{sec:model}
\subsection{System under consideration}
\begin{figure}
\includegraphics[width=\linewidth]{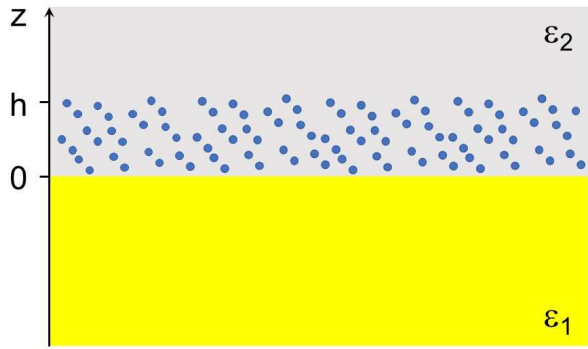}
\caption{\label{fig:sketch} The sketch of the structure under consideration.}
\end{figure}

Let us consider an interface between a semi-infinite metal with the dielectric function $\epsilon_1(\omega)$ and a semi-infinite dielectric with the dielectric function $\epsilon_2$ and direct the $z$ coordinate axis along the normal to the interface from the metal to the dielectric. Let us assume that active centers are distributed within the dielectric from $z=0$ to $z=h$ (see Fig. \ref{fig:sketch}). We accept the Drude model for the dielectric function of the metal,
\begin{equation}\label{eq:drude}
\epsilon_1(\omega)=\epsilon_{\infty}-\frac{\omega_p^2}{\omega(\omega+i\Gamma)},
\end{equation}
where $\epsilon_{\infty}$ is the offset originating from the interband transitions, $\omega_p$ is the metal plasma frequency and $\Gamma$ is the relaxation constant.\\
\subsection{Evolution of the active centers polarization}
The radiation emitted from the system can be regarded as being originating from the active centers dipoles ${\bf p}({\bf r},t)$, ${\bf r}$ being the radius vector of the center, oscillating near the active center transition frequency. The electromagnetic field which exists in the system can be decomposed into the negative and positive frequency parts as
\begin{equation}
{\bf E}(t)=\frac{1}{2}\left[{\bf E}^{(-)}(t)e^{i\omega t}+{\bf E}^{(+)}(t)e^{-i\omega t}\right]
\end{equation}
with
\begin{equation}
{\bf E}^{(\pm)}(t)=[{\bf E}^{(\mp)}(t)]^*
\end{equation}
being the slowly varying amplitudes in which we have omitted the dependence on the active center coordinates for the sake of brevity.\\
The polarization ${\bf P}({\bf r},t)=N{\bf p}({\bf r},t)$ with $N$ being the number density of active centers takes a similar form,
\begin{equation}
{\bf P}(t)=\frac{1}{2}\left[{\bf P}^{(-)}(t)e^{i\omega t}+{\bf P}^{(+)}(t)e^{-i\omega t}\right],
\end{equation}
where the slow amplitudes ${\bf P}^{(\pm)}(t)$ satisfy the optical Bloch equations \cite{Haken}
\begin{eqnarray}\label{eq:bloch1}
\frac{\partial {\bf P}^{(+)}(t)}{\partial t}=-(\gamma_{\perp }-i\Delta){\bf P}^{(+)}(t)\nonumber\\
-\frac{i}{3\hbar } \mu^2D(t){\bf E}^{(+)}(t),
\end{eqnarray}
\begin{eqnarray}\label{eq:bloch2}
\frac{\partial D(t)}{\partial t}=-\gamma_{\parallel }\left[D(t)-D_0\right]\nonumber\\
+\frac{i}{2\hbar }\left[{\bf P}^{(-)}(t){\bf E}^{(+)}(t)-{\bf P}^{(+)}(t){\bf E}^{(-)}(t)\right].
\end{eqnarray}
Here $\Delta=\omega-\omega_0$ is the resonance detuning, $\gamma_{\perp}$ and $\gamma_{\parallel}$ are the transverse (phase) and longitudinal (energy) relaxation rates of the active center transition, respectively, $\mu$ is the transition dipole moment, $D=Nw$, $w$ is the population inversion between the upper and lower states and $D_0$ is the equilibrium value of $D$ determined by the optical pumping through the other quantum levels of the active centers. The factor $1/3$ in front of $\mu^2$ in Eq. (\ref{eq:bloch1}) originates from the averaging over the dipole orientations. In a close vicinity to the surface the relaxation rate $\gamma_{\parallel}$ depends on the $z$-coordinate \cite{Chance74}.\\
The field amplitudes ${\bf E}^{(\pm)}({\bf r},t)$ in the above equations should be regarded as the amplitudes of the local field in the gain medium which depends itself on its polarization. This dependence can be written in the form of an integral equation
\begin{equation}\label{eq:integral}
{\bf E}^{(+)}({\bf r},t)=\int_V^{\prime} \bar{\bf F}({\bf r},{\bf r}^{\prime}){\bf P}^{(+)}({\bf r}^{\prime},t)d{\bf r}^{\prime},
\end{equation}
where $V$ is the volume of the gain medium and the prime above the integral sign implies removal of the point ${\bf r}^{\prime}={\bf r}$ from the integration. The kernel in this equation, $\bar{\bf F}({\bf r},{\bf r}^{\prime})$, is the so-called field susceptibility tensor which relates the electromagnetic field at the point ${\bf r}$ generated by a classical dipole with the dipole moment itself located at the point ${\bf r}^{\prime}$ \cite{Sipe84}.\\
In the bulk of the medium, Eq. (\ref{eq:integral}) is reduced to the Lorentz relation
\begin{equation}\label{eq:lorentz}
{\bf E}^{(+)}({\bf r},t)=\frac{4\pi}{3\epsilon_2}{\bf P}^{(+)}({\bf r},t).
\end{equation}
However at distances of the order of the wavelength or less from an interface the Lorentz field should be complemented by the dipole field reflected from the interface that can be written as
\begin{equation}
{\bf E}^{(+)}({\bf r},t)=\frac{4\pi}{3\epsilon_2}{\bf P}^{(+)}({\bf r},t)+\int_V \bar{\bf F}^R({\bf r},{\bf r}^{\prime}){\bf P}^{(+)}({\bf r}^{\prime},t)d{\bf r}^{\prime},
\end{equation}
where the superscript $R$ denotes the reflected field contribution.\\
\section{Lasing condition}\label{sec:lasing}
\subsection{Instability in the system}
Let us assume now that initially at $t=0$ there is no electromagnetic field oscillating at the frequency $\omega$. We investigate the evolution of the system when there emerges a field of an infinitesimal amplitude $\delta {\bf E}({\bf r},t)$. \\
In the linear approximation in $\delta {\bf E}({\bf r},t)$ one obtains the equations
\begin{eqnarray}\label{eq:linear}
\frac{\partial {\bf P}^{(+)}({\bf r},t)}{\partial t}=-(\gamma_{\perp }-i\Delta){\bf P}^{(+)}({\bf r},t)\nonumber\\
-\frac{i}{3\hbar } \mu^2D_0{\bf E}^{(+)}({\bf r},t)
\end{eqnarray}
and
\begin{eqnarray}
{\bf E}^{(+)}({\bf r},t)=\delta{\bf E}^{(+)}({\bf r},t)+\frac{4\pi}{3\epsilon_2}{\bf P}^{(+)}({\bf r},t)\nonumber\\
+\int_V \bar{\bf F}^R({\bf r},{\bf r}^{\prime}){\bf P}^{(+)}({\bf r}^{\prime},t)d{\bf r}^{\prime}.
\end{eqnarray}
Introducing the spatial Fourier transforms and the Laplace transforms in time,
\begin{equation}\label{eq:Efourier}
{\bf E}^{(+)}
({\bf k}_{\parallel},s;z)=\int\int_0^{\infty} {\bf E}^{(+)}({\bf r},t)e^{-i{\bf k}_{\parallel}\cdot {\bf r}_{\parallel}}e^{-st}dtd{\bf r}_{\parallel},
\end{equation}
\begin{equation}\label{eq:Pfourier}
{\bf P}^{(+)}({\bf k}_{\parallel},s;z)=\int\int_0^{\infty} {\bf P}^{(+)}({\bf r},t)e^{-i{\bf k}_{\parallel}\cdot {\bf r}_{\parallel}}e^{-st}dtd{\bf r}_{\parallel}
\end{equation}
with ${\bf r}_{\parallel}$ and ${\bf k}_{\parallel}$ being the radius vector and the wave vector components along the interface, respectively, one comes to the integral equation
\begin{eqnarray}\label{eq:integralFL}
\int_0^h\bar{\bf F}^R({\bf k}_{\parallel};z,z^{\prime}){\bf P}^{(+)}({\bf k}_{\parallel},s;z^{\prime})dz^{\prime}\nonumber\\
-\nu(s){\bf P}^{(+)}({\bf k}_{\parallel},s;z)=-\delta{\bf E}^{(+)}({\bf k}_{\parallel},s;z),
\end{eqnarray}
where 
\begin{equation}
\nu(s)=\frac{1-(4\pi/3\epsilon_2)\chi(s)}{\chi(s)}
\end{equation}
with
\begin{equation}
\chi(s)=-\frac{i}{3\hbar}\frac{\mu^2D_0}{s+\gamma_{\perp}-i\Delta},
\end{equation}
the quantity $\bar{\bf F}^R({\bf k}_{\parallel};z,z^{\prime})$ is determined by the Fourier transform
\begin{equation}\label{eq:Ffourier}
\bar{\bf F}^R({\bf r},{\bf r}^{\prime})=\frac{1}{(2\pi)^2}\int \bar{\bf F}^R({\bf k}_{\parallel};z,z^{\prime})e^{i{\bf k}_{\parallel}\cdot ({\bf r}_{\parallel}-{\bf r}^{\prime}_{\parallel})}d{\bf k}_{\parallel}
\end{equation}
and $\delta{\bf E}^{(+)}({\bf k}_{\parallel},s;z)$ is the spatial Fourier transform and the Laplace transform in time of the quantity  $\delta {\bf E}^{(+)}({\bf r},t)$.\\
The explicit form of the tensor $\bar{\bf F}^R({\bf k}_{\parallel};z,z^{\prime})$ is found in Ref. \cite{Sipe84} and can be written as
\begin{equation}\label{eq:FR}
\bar{\bf F}^R({\bf k}_{\parallel};z,z^{\prime})=\bar{\bf f}({\bf k}_{\parallel})e^{iq_2(k_{\parallel})(z+z^{\prime})}
\end{equation}
with 
\begin{equation}
q_2(k_{\parallel})=\sqrt{\frac{\omega^2}{c^2}\epsilon_2-k_{\parallel}^2},
\end{equation}
where the quantity $\bar{\bf f}({\bf k}_{\parallel})$ is given in Appendix A. This allows one to rewrite Eq. (\ref{eq:integralFL}) as follows
\begin{equation}\label{eq:matrix}
S(k_{\parallel})\hat{M}(k_{\parallel},s)\vec{P}(k_{\parallel},s)=-\delta\vec{E}(k_{\parallel},s),
\end{equation}
where
\begin{equation}
\vec{P}(k_{\parallel},s)=\int_0^h {\bf P}^{(+)}(k_{\parallel},s;z)e^{iq_2(k_{\parallel})z}dz,
\end{equation}
\begin{equation}
\delta\vec{E}(k_{\parallel},s)=\int_0^h \delta\mathbf{E}^{(+)}(k_{\parallel},s;z)e^{iq_2(k_{\parallel})z}dz,
\end{equation}
\begin{equation}
S(k_{\parallel})=\int_0^h e^{2iq_2(k_{\parallel})z}dz=\frac{1}{2iq_2(k_{\parallel})}\left[e^{2iq_2(k_{\parallel})h}-1\right]
\end{equation}
and the matrix $\hat{M}$ is defined as
\begin{equation}
\hat{M}(k_{\parallel},s)=\hat{f}(k_{\parallel})-\frac{\nu(s)}{S(k_{\parallel})}\hat{I}
\end{equation}
with $\hat{f}$ being the matrix of the tensor $\bar{\bf f}$ and $\hat{I}$ is the unit $3\times 3$ matrix. We have also taken into account that because of the axial symmetry of the problem nothing depends on the direction of the vector ${\bf k}_{\parallel}$.\\
Equation (\ref{eq:matrix}) describes the evolution of the active centers polarization associated with the spatial Fourier component specified by $k_{\parallel}$. All the quantities here are averaged over the gain medium thickness with the account of the reflected electromagnetic field phase factor. The function $S(k_{\parallel})$ appears because the downward propagating waves are reflected back upward \cite{Sipe84}. \\
As it follows from Eq. (\ref{eq:matrix}), the evolution of the polarization is determined by the zeros of the determinant of the matrix $\hat{M}(k_{\parallel},s)$ which provide the poles of the polarization Laplace transform. On the other hand, these zeros are related with the eigenvalues  $\phi_j(k_{\parallel})$ of the matrix $\hat{f}(k_{\parallel})$ through the equation
\begin{equation}\label{eq:phi}
\phi_j(k_{\parallel})=\frac{\nu(s_j)}{S(k_{\parallel})},
\end{equation}
which implicitly determines the poles of the Laplace transform, $s_j(k_{\parallel})$. Resolving Eq. (\ref{eq:phi}) relatively $s_j$ one finds
\begin{equation}
s_j(k_{\parallel})=-\gamma_{\perp}+i\Delta^{\prime}+\frac{2\pi}{3\hbar}\mu^2D_0\mathcal{F}_j(k_{\parallel}),
\end{equation}
where
\begin{equation}
\Delta^{\prime}=\Delta-\frac{4\pi}{9\hbar\epsilon_2}\mu^2D_0,
\end{equation}
\begin{equation}
\mathcal{F}_j(k_{\parallel})=S(k_{\parallel})\tilde{\phi}_j(k_{\parallel})
\end{equation}
and the quantities $\tilde{\phi}_j(k_{\parallel})$ for $s$ and $p$ polarizations are given in Appendix A.\\ 
Let us note that the quantities $\tilde{\phi}_j(k_{\parallel})$ are determined by the reflection coefficients from the metal surface, whereas the function $S(k_{\parallel})$ represents an averaged phase factor which the dipole field acquires when propagating downward and being reflected back upward. This means that the quantities $\mathcal{F}_j(k_{\parallel})$ characterize the feedback loop provided by the reflecting surface.\\ 
Now separating the real and imaginary parts of the poles, $s_j(k_{\parallel})=\sigma_j(k_{\parallel})+i\omega_j(k_{\parallel})$, one obtains
\begin{equation}\label{eq:sigma}
\sigma_j(k_{\parallel})=-\gamma_{\perp}+\frac{2\pi}{3\hbar}\mu^2D_0\text{Re}\mathcal{F}_j(k_{\parallel})
\end{equation}
and
\begin{equation}\label{eq:omega}
\omega_j(k_{\parallel})=\Delta^{\prime}+\frac{2\pi}{3\hbar}\mu^2D_0\text{Im}\mathcal{F}_j(k_{\parallel}).
\end{equation}
The inverse Laplace transform indicates that for every spatial mode specified by $k_{\parallel}$ the polarization amplitude of the active centers, and consequently the emitted field amplitude, vary with time as $\exp[\sigma_j(k_{\parallel})t]$. The field generation (lasing) implies an exponential growth of the amplitude which occurs if $\sigma_j(k_{\parallel})>0$. The latter quantity, Eq. (\ref{eq:sigma}), contains two contributions: the phase relaxation rate with minus sign, which describes the decay of the polarization, and the second term, which stems from the surface feedback. Lasing is only possible if the latter term is positive and exceeds in magnitude the first one, i.e.
\begin{equation}\label{eq:threshold}
\frac{2\pi}{3\hbar}\mu^2D_0\text{Re}\mathcal{F}_j(k_{\parallel})>\gamma_{\perp}.
\end{equation}
This can be interpreted as a condition that the gain supported by the feedback loop prevails over the losses in the polarization of the active centers.\\
The positiveness of the left-hand side in Eq. (\ref{eq:threshold}) can be realized in two cases. One of them, where both $D_0$ and $\text{Re}\mathcal{F}_j(k_{\parallel})$ are positive, corresponds to the population inversion at the transition and can be referred to as "conventional lasing". The other possibility takes place when both $D_0$ and $\text{Re}\mathcal{F}_j(k_{\parallel})$ are negative and can be therefore termed "lasing without inversion". Different signs of $\text{Re}\mathcal{F}_j(k_{\parallel})$ indicate different character of the feedback loop. In the first case the feedback is negative, and the gain is provided by the population inversion. In the second case the feedback from the surface is positive and the inversion is not necessary for the gain.\\
The imaginary part of the pole, Eq. (\ref{eq:omega}), determines the frequency of the generated field (the frequency pulling effect), 
\begin{equation}\label{eq:pulling}
\omega_g=\omega_0+\frac{4\pi}{9\hbar\epsilon_2}\mu^2D_0-\frac{2\pi}{3\hbar}\mu^2D_0\text{Im}\mathcal{F}_j(k_{\parallel}).
\end{equation}
Let us note that assuming a slow variation of the amplitude ${\bf P}^{(+)}(t)$ in time we implied a small detuning $\Delta$. Therefore the applicability criterion of Eq. (\ref{eq:pulling}) is $\mid \omega_g-\omega_0\mid\ll \omega_0$ that requires
\begin{equation}
\frac{2\pi}{3\hbar}\mu^2D_0\left\vert\text{Im}\mathcal{F}_j(k_{\parallel})\right\vert\ll\omega_0.
\end{equation}
The latter condition imposes also an upper limit for the left-hand side part of Eq. (\ref{eq:threshold}). This implies, in particular, that the developed theory is not applicable very close to the grazing propagation where the quantities $\tilde{\phi}_j(k_{\parallel})$ take very large values.\\
The lasing condition, Eq. (\ref{eq:threshold}), involves a number of independent parameters among which $\mu$, $D_0$ and $\gamma_{\perp}$ refer to the active centers and $\mathcal{F}_j(k_{\parallel})$ is determined by the feedback from the surface. For a uniform characterization of lasing it is convenient to introduce a dimensionless parameter related to the active centers as follows
\begin{equation}
\eta=\frac{2\pi\mu^2D_0}{3\hbar\gamma_{\perp}}.
\end{equation}
In such a description, the dimensionless parameter 
\begin{equation}
\eta_{th}=\frac{1}{\text{Re}\mathcal{F}_j(k_{\parallel})},
\end{equation}
specifies the threshold value of $\eta$ for given polarization and $k_{\parallel}$ above which lasing becomes possible. Then the lasing condition takes the form
\begin{equation}\label{eq:lasing}
\eta>\eta_{th}
\end{equation}
for conventional lasing and 
\begin{equation}\label{eq:lwi}
\vert\eta\vert>\vert\eta_{th}\vert
\end{equation}
for lasing without inversion.\\
The lasing condition considered above refers to the initial stage of the laser generation where the generated field is relatively weak. When the field amplitude in the system becomes comparable with the saturation field of the transition one must consider the deviation of the population difference density $D$ from its equilibrium value $D_0$ which is described by the second Bloch equation (\ref{eq:bloch2}). The relevant derivation can be done in a standard way (see, e.g. Refs. \cite{Bordo13} and \cite{Bordo17} for the detail) and is beyond the scope of the present paper. As a result, the gain gets saturated and exactly compensates the losses, that establishes the steady state regime and builds up coherence.\\
\subsection{Energy balance in the system}
The criterion of generation can be alternatively obtained from Poynting's theorem written for the electromagnetic field in the gain medium (in Gaussian units) \cite{Stratton41},
\begin{eqnarray}
\frac{c}{4\pi}\int_S ({\bf E}\times{\bf H}){\bf n}da+\int_V{\bf E}{\bf J}dv\nonumber\\
=-\frac{1}{4\pi}\int_V\left({\bf E}\frac{\partial\bf D}{\partial t}+{\bf H}\frac{\partial{\bf B}}{\partial t}\right)dv,
\end{eqnarray}
where  ${\bf D}=\epsilon_2{\bf E}$ and we have used the standard notations. The right-hand side part of this equation represents the rate of change of the electromagnetic energy, $\partial W_{em}/\partial t$, contained in the volume $V$ of the gain medium taken with the minus sign. The first term in the left-hand side is the flux of the electromagnetic energy across the surface $S$ which embraces the volume $V$, while the second term expresses the rate of doing work on driving the displacement current, 
\begin{equation}
{\bf J}=\frac{\partial{\bf P}}{\partial t},
\end{equation}
by the electromagnetic field.\\
In the resonant approximation ($\omega\approx\omega_0$), Eq. (\ref{eq:linear}) along with its complex conjugate can be reduced to the damped harmonic oscillator equation (see also Ref. \cite{Pantel})
\begin{equation}
{\ddot{\bf P}}+2\gamma_{\perp}{\dot{\bf P}}+\omega^2{\bf P}=a{\bf E},
\end{equation}
where $a=-2\omega\mu^2D_0/3\hbar$ and the dot above a symbol denotes the time derivative. Now the term representing the rate of doing work on driving the current ${\bf J}$ can be expressed as
\begin{equation}\label{eq:work}
\int_V {\bf E}{\dot{\bf P}}dv=\text{sgn} (a)\frac{\partial W_m}{\partial t}+\frac{2\gamma_{\perp}}{a}\int_V{\dot{\bf P}}^2dv,
\end{equation}
where $\text{sgn}(a)$ is the signum function and the quantity
\begin{equation}
W_m=\frac{1}{2\vert a\vert}\int_V\left({\dot{\bf P}}^2+\omega^2{\bf P}^2\right)dv
\end{equation}
has a sense of the mechanical (kinetic plus potential) energy accumulated in the active centers \cite{Jackson}. The other term in the right-hand side of Eq. (\ref{eq:work}) provides the decay rate of the mechanical energy due to the phase relaxation of the polarization. \\
Taking the time average over a period of the field oscillations, which is denoted by the angle brackets, one obtains that $\langle \partial W_{em}/\partial t\rangle =0$, i.e. the total electromagnetic energy in the system is not changed for a period as it is {\it a priori} expected (see Ref. \cite{Bordo16}, Appendix C for the detail).\\ 
The time averaging of Eq. (\ref{eq:work}) leads to the equation 
\begin{eqnarray}\label{eq:mech_balance}
\text{sgn}(D_0)\Big\langle\frac{\partial W_m}{\partial t}\Big\rangle
\approx \frac{\omega}{2}\frac{h}{(2\pi)^2}\nonumber\\
\times\int \sum_i\text{Im}\left[\phi_i(k_{\parallel})S(k_{\parallel})
-i\frac{2\pi}{\eta}\right] \vert\tilde{P}_i^{(+)}(k_{\parallel};t)\vert^2d{\bf k}_{\parallel},\nonumber\\
\end{eqnarray}
which implies that the mechanical energy accumulated in the active centers and associated with certain polarization and the wavevector $k_{\parallel}$ gradually increases after every period of oscillations if either Eq. (\ref{eq:lasing}) or Eq. (\ref{eq:lwi}) is fulfilled (see Appendix B).\\
Equation (\ref{eq:mech_balance}) shows how different electromagnetic modes specified by $k_{\parallel}$ contribute to the generation process. In the course of the instability development, the contributions from $k_{\parallel}$ for which the lasing condition is not fulfilled vanish due to the exponential decay of the amplitudes $\vert\tilde{P}_i^{(+)}(k_{\parallel};t)\vert$ with time. On the other hand, the modes for which the lasing condition holds grow exponentially provided that the seed field $\delta {\bf E}({\bf r},t)$ has non-zero spatial Fourier components at the relevant $k_{\parallel}$. \\
\subsection{Lasing in an optically thick gain medium}
It is interesting to follow how the above derived lasing conditions are transformed in an optically thick gain medium where $(\omega/c)h\gg 1$. One can see that the mechanical energy increase in Eq. (\ref{eq:mech_balance}) is represented in terms of an integral over the wave vectors $k_{\parallel}$. The contribution of the rapidly oscillating exponent $\exp[2iq_2(k_{\parallel})h]$ in the function $S(k_{\parallel})$ to the integral is negligible and one can adopt the approximation
\begin{equation}\label{eq:approx}
S(k_{\parallel})\approx -\frac{1}{2iq_2(k_{\parallel})}.
\end{equation}
This eliminates the dependence of the lasing condition on the gain medium thickness.\\
Alternatively, this can be seen from Eq. (\ref{eq:B5}) where the integral term represents the contribution of the reflected field in the polarization Fourier transform evolution. Considering Eqs. (\ref{eq:B6}) and (\ref{eq:FR}) one concludes that for the points distant from the surface such that $z\gg c/\omega$ the Fourier transform rapidly oscillates with $k_{\parallel}$ that gives a negligible contribution to the reflected field itself. The physical sense of this result is as follows. The backaction of the surface on the ensemble of dipoles decays over a distance of the order of the wavelength and the further increase of the gain medium thickness does not influence the lasing condition.\\
Taking into account the approximation (\ref{eq:approx}) one finds
\begin{equation}\label{eq:Fs}
\text{Re}\mathcal{F}_s(k_{\parallel})=\frac{1}{\vert\epsilon_1\vert + \epsilon_2}\text{Im}\frac{q_1(k_{\parallel})}{q_2(k_{\parallel})}
\end{equation}
and
\begin{equation}\label{eq:Fp}
\text{Re}\mathcal{F}_p(k_{\parallel})=\frac{q_2^2(k_{\parallel})-k_{\parallel}^2}{2\epsilon_2q_2^2(k_{\parallel})}\text{Im}\frac{\epsilon_2q_1(k_{\parallel})+\vert\epsilon_1\vert q_2(k_{\parallel})}{\epsilon_2q_1(k_{\parallel})-\vert\epsilon_1\vert q_2(k_{\parallel})} 
\end{equation}
for $s$ and $p$ polarizations of the active centers dipoles, respectively, where we have taken into account that $\epsilon_1$ is a real negative quantity in neglect of its imaginary part. Both quantities given by Eqs. (\ref{eq:Fs}) and (\ref{eq:Fp}) are non-zero only if $q_2(k_{\parallel})$ is real, i.e. $k_{\parallel}<(\omega/c)\sqrt{\epsilon_2}$. This means that lasing is possible for the electromagnetic waves propagating in the gain medium and impossible for evanescent (non-radiative) waves propagating along the metal/gain medium interface.\\
\section{Numerical results}\label{sec:numerical}
We illustrate the above developed theory by some numerical calculations performed for a structure which simulates the one used in the experiments on SPP stimulated emission at telecom wavelength (1532 nm) at a gold surface with erbium doped glass as a gain medium \cite{Zhang08}. The calculations have been carried out for the following values of parameters: $\epsilon_{\infty}=9$, $\omega_p=13.8\times 10^{15}$ s$^{-1}$, $\Gamma=0.11\times 10^{15}$ s$^{-1}$ \cite{Shalaev10} and $\epsilon_2=1.5^2$. For Er$^{3+}$ ions one has $\mu=3.50\times 10^{-32}$ Cm $= 0.01$ D and $\gamma_{\perp}=1/T_2$ with $T_2=80$ $\mu$s \cite{Sellars09}. The number density can attain the value $N=5.3\times 10^{20}$ cm$^{-3}$ \cite{Yan97}. Assuming that the population inversion is maximal ($w=1$), one obtains $\eta\approx 8.9\times 10^3$. This value is, however, an upper limit which can be reached at cryogenic temperatures and the possible values of $\eta$ at higher temperatures are somewhat lower.\\
\begin{figure}
\includegraphics[width=\linewidth]{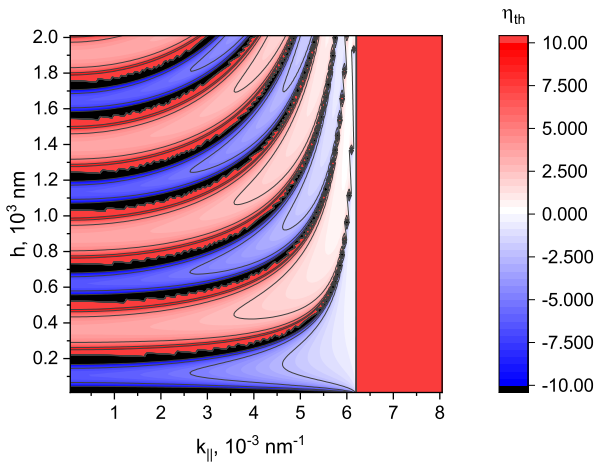}
\caption{\label{fig:s} The map of the lasing threshold, $\eta_{th}$, versus the wavevector parallel to the interface and the gain medium thickness for the generation of $s$-polarized electromagnetic field at 1542 nm.}
\end{figure}
\begin{figure}
\includegraphics[width=\linewidth]{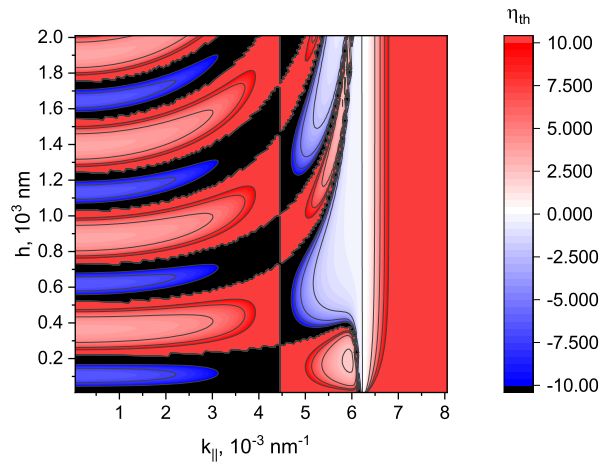}
\caption{\label{fig:p} Same as in Fig. \ref{fig:s}, but for $p$-polarized electromagnetic field.}
\end{figure}
Figure \ref{fig:s} demonstrates the lasing threshold parameter plotted versus $k_{\parallel}$ and $h$. The levels of $\eta_{th}$ are only shown in the range between $-10$ and $10$, where the threshold is lowest. The red regions correspond to conventional lasing, whereas the blue ones correspond to lasing without inversion. The lighter tones indicate lower lasing thresholds. One can see that for a fixed value of $k_{\parallel}$ the character of lasing varies periodically with $h$. The lasing threshold decreases when $k_{\parallel}$ approaches the domain of grazing propagation in the gain medium where $k_{\parallel}\approx (\omega/c)\sqrt{\epsilon_2}\approx 6.1\times 10^{-3}$ nm$^{-1}$. Behind this value the non-radiative region is disposed where lasing is impossible.\\
Somewhat similar one observes for lasing in $p$ polarization (see Fig. \ref{fig:p}). However this time the character of lasing switches at $k_{\parallel}=(\omega/c)\sqrt{\epsilon_2/2}\approx 4.4\times 10^{-3}$ nm$^{-1}$ where $\tilde{\phi}_{p2}$ changes its sign. Besides that, conventional lasing is possible in the non-radiative region just behind the grazing propagation domain that corresponds to SPP generation.\\
\section{Conclusion}\label{sec:conclusion}
In this paper, we have considered lasing in an optically thin gain medium in contact with a metal surface. We have derived the lasing condition for such a structure basing on a rigorous account of the local field in the gain layer. \\
We have shown that the lasing threshold is determined by a product of the population inversion in active centers with the factor which originates from the backaction from the metal surface and depends on the gain medium thickness. Such a condition predicts, besides conventional lasing, lasing without inversion if the latter factor is negative. For a given wave vector component parallel to the interface, this factor is a periodic function of the gain layer thickness that suggests a significant reduction of the overall structure size without loss in laser functionality. This effect disappears in an optically thick gain medium.\\
The predicted possibility of lasing without inversion opens up a new avenue for the development of nanoscale coherent light sources which will not require powerful pumping and will be therefore low-cost energy-saving nanodevices. Another possible application of this effect is amplification of very weak nano-localized electromagnetic fields that can be exploited for ultra-sensitive surface chemical analysis. The findings of the present paper can stimulate further investigations in this promising direction.\\
\section*{Acknowledgments}
This research is kindly supported by the IE-Industrial Elektronik project (SFD-17-0036) which has received EU co-financing from the European Social Fund.\\
\appendix
\section{Field susceptibility tensor and its eigenvalues}
The tensor $\bar{{\bf f}}({\bf k}_{\parallel})$ introduced by Eq. (\ref{eq:FR}) has the form (see Ref. \cite{Sipe84})
\begin{equation}
\bar{{\bf f}}({\bf k}_{\parallel})=2\pi i\frac{\tilde{\omega}^2}{q_2(k_{\parallel})}\left[\hat{s}\hat{s}R^s(k_{\parallel})+\hat{p}_+\hat{p}_-R^p(k_{\parallel})\right],
\end{equation}
where $\tilde{\omega}=\omega/c$, $\hat{s}=\hat{k}_{\parallel}\times\hat{z}$ with $\hat{k}_{\parallel}$ and $\hat{z}$ being the unit vectors oriented along the corresponding directions,
\begin{equation}
\hat{p}_{\pm}=\left(\tilde{\omega}\sqrt{\epsilon_2}\right)^{-1}\left[k_{\parallel}\hat{z}\mp q_2(k_{\parallel})\hat{k}_{\parallel}\right],
\end{equation}
\begin{equation}
R^s(k_{\parallel})=\frac{q_2(k_{\parallel})-q_1(k_{\parallel})}{q_2(k_{\parallel})+q_1(k_{\parallel})}
\end{equation}
and
\begin{equation}
R^p(k_{\parallel})=\frac{\epsilon_1 q_2(k_{\parallel})-\epsilon_2 q_1(k_{\parallel})}{\epsilon_1 q_2(k_{\parallel})+\epsilon_2 q_1(k_{\parallel})}
\end{equation}
with $q_1(k_{\parallel})=\sqrt{(\omega/c)^2\epsilon_1-k_{\parallel}^2}$ are the Fresnel reflection coefficients for $s$ and $p$ polarizations, respectively.\\
The matrix of the tensor $\bar{{\bf f}}({\bf k}_{\parallel})$ in the basis of the vectors $\hat{s}$, $\hat{k}_{\parallel}$ and $\hat{z}$ takes the following form:
\begin{equation}
\hat{f}=2\pi i\left(\matrix{\frac{\tilde{\omega}^2}{q_2}R^s & 0 & 0 \cr
0 & -\frac{q_2}{\epsilon_2}R^p & -\frac{k_{\parallel}}{\epsilon_2}R^p \cr
0 & \frac{k_{\parallel}}{\epsilon_2}R^p & \frac{k_{\parallel}^2}{\epsilon_2q_2}R^p }\right).
\end{equation}
Correspondingly, its eigenvalues are found as $\phi_j(k_{\parallel})=2\pi i \tilde{\phi}_j(k_{\parallel})$ with
\begin{equation}\label{eq:phis}
\tilde{\phi}_s(k_{\parallel})=\frac{\tilde{\omega}^2}{q_2(k_{\parallel})}R^s(k_{\parallel}),
\end{equation}
\begin{equation}
\tilde{\phi}_{p1}(k_{\parallel})=0
\end{equation}
and 
\begin{equation}\label{eq:phip}
\tilde{\phi}_{p2}(k_{\parallel})=\frac{k_{\parallel}^2-q_2^2(k_{\parallel})}{\epsilon_2q_2(k_{\parallel})}R^p(k_{\parallel}).
\end{equation}
\section{Poynting's theorem}
The time-averaged quantities which enter Poynting's theorem are found as 
\begin{equation}
\frac{c}{4\pi}\Big\langle \int_S ({\bf E}\times{\bf H}){\bf n}da \Big\rangle = \frac{\omega}{2}\int_V\text{Im}\left({\bf E}^{(+)}{\bf P}^{(-)}\right)dv
\end{equation}
and
\begin{equation}
\frac{2\gamma_{\perp}}{a}\Big\langle \int_V{\dot{\bf P}}^2dv \Big\rangle = \frac{\gamma_{\perp}\omega^2}{a}\int_V\vert{\bf P}^{(+)}\vert^2dv.
\end{equation}
As a result, the balance of energy for a period takes the form
\begin{widetext}
\begin{equation}\label{eq:period}
\text{sgn}(D_0)\Big\langle\frac{\partial W_m}{\partial t}\Big\rangle
\approx \frac{\omega}{2}\left[\text{Im}\sum_{i,j}\int_V\int_VF^R_{ij}({\bf r},{\bf r}^{\prime})P_j^{(+)}({\bf r}^{\prime},t)P_i^{(-)}({\bf r},t)dvdv^{\prime}
+\frac{2\gamma_{\perp}\omega}{a}\sum_i\int_VP_i^{(+)}({\bf r},t)P_i^{(-)}({\bf r},t)dv\right],
\end{equation}
where the summation runs over the cartesian components and we have neglected the small quantity $\delta{\bf E}^{(+)}$.\\ 
Let us introduce here the Fourier transforms
\begin{equation}
P_i^{(+)}({\bf r},t)=\frac{1}{(2\pi)^2}\int P_i^{(+)}(k_{\parallel};z,t)e^{i{\bf k}_{\parallel}\cdot{\bf r}_{\parallel}}d{\bf k}_{\parallel},
\end{equation}
where the quantities $P_i^{(+)}(k_{\parallel};z,t)$ in the adopted approximation ($\Delta\approx 0$, $\delta{\bf E}^{(+)}\approx 0$) satisfy the equations [see Eq. (\ref{eq:linear})]
\begin{equation}\label{eq:B5}
\frac{\partial P_i^{(+)}(k_{\parallel};z,t)}{\partial t}\approx -\gamma_{\perp } P_i^{(+)}(k_{\parallel};z,t)
-\frac{i}{3\hbar } \mu^2D_0\left[\frac{4\pi}{3\epsilon_2}P_i^{(+)}(k_{\parallel};z,t)
+\int_0^h \sum_j F_{ij}^R(k_{\parallel};z,z^{\prime})P_j^{(+)}(k_{\parallel};z^{\prime},t)dz^{\prime}\right].
\end{equation}
Taking into account the explicit form of the tensor $\bar{\bf F}^R({\bf k}_{\parallel};z,z^{\prime})$, Eq. (\ref{eq:FR}), one concludes that the above equations allow the solutions of the form
\begin{equation}\label{eq:B6}
P_i^{(+)}(k_{\parallel};z,t)=P_i^{(+)}(k_{\parallel};t)e^{iq_2(k_{\parallel})z}.
\end{equation}
Then one obtains Eq. (\ref{eq:period}) as follows
\begin{equation}\label{eq:hermitian}
\text{sgn}(D_0)\Big\langle\frac{\partial W_m}{\partial t}\Big\rangle
\approx \frac{\omega}{2}\frac{h}{(2\pi)^2}
\int \text{Im}\left\{\sum_{i,j}\left[f_{ij}(k_{\parallel})S(k_{\parallel})
+i\frac{2\gamma_{\perp}\omega}{a}\delta_{ij}\right] P_j^{(+)}(k_{\parallel};t)P_i^{(+)*}(k_{\parallel};t)\right\}d{\bf k}_{\parallel},
\end{equation}
where $\delta_{ij}$ is the Kronecker delta.\\
The Hermitian form which enters the integrand in Eq. (\ref{eq:hermitian}) can be reduced to its normal form with the use of the linear transformation $P_i^{(+)}=\sum_j T_{ij}\tilde{P}_j^{(+)}$ that results in the equation
\begin{equation}
\text{sgn}(D_0)\Big\langle\frac{\partial W_m}{\partial t}\Big\rangle
\approx \frac{\omega}{2}\frac{h}{(2\pi)^2}
\int \sum_i\text{Im}\left[\phi_i(k_{\parallel})S(k_{\parallel})
-i\frac{2\pi}{\eta}\right] \vert\tilde{P}_i^{(+)}(k_{\parallel};t)\vert^2d{\bf k}_{\parallel},
\end{equation}
\end{widetext}
where $\phi_i(k_{\parallel})=2\pi i\tilde{\phi}_i(k_{\parallel})$ are the eigenvalues of the matrix $f_{ij}(k_{\parallel})$ (see Appendix A) and we have used the definitions of the parameters $a$ and $\eta$. From here the condition that the mechanical energy, associated with certain polarization ($s$ or $p$) and the wavevector $k_{\parallel}$, accumulated in the active centers will gradually increase after every period can be written as
\begin{equation}
\text{Re}\left[\tilde{\phi}_i(k_{\parallel})S(k_{\parallel})\right]-\frac{1}{\eta}>0
\end{equation}
if $D_0$ is positive and as
\begin{equation}
\text{Re}\left[\tilde{\phi}_i(k_{\parallel})S(k_{\parallel})\right]-\frac{1}{\eta}<0
\end{equation}
if $D_0$ is negative that is equivalent to the lasing conditions, Eqs. (\ref{eq:lasing}) and (\ref{eq:lwi}), respectively.\\

\end{document}